\documentclass[fleqn,10pt, onecolumn]{wlscirep}
\usepackage[utf8]{inputenc}
\usepackage[T1]{fontenc}

\usepackage{url}

\usepackage{hyperref}

\title{Niche to normality – an interdisciplinary review of Vehicle-to-Grid}

\author[1,+,*]{Bj{\"o}rn C. P. Sturmberg}
\author[1,+]{Laura Jones}
\author[1,+]{Kathryn Lucas-Healey}
\author[1]{Monirul Islam}
\author[1]{Hugo Temby}
\affil[1]{Research School of Electrical, Energy and Materials Engineering, The Australian National University, Canberra, 2601, Australia}

\affil[*]{bjorn.sturmberg@anu.edu.au}

\affil[+]{these authors contributed equally to this work}

\begin{abstract}
Vehicle-to-Grid (V2G) capabilities, which enable electric vehicles to discharge power from their batteries for external uses, epitomise the coupling of the electricity and transport sectors. To thrive at the nexus of these large and well-established sectors V2G services must deliver technical, economic and social values to many stakeholders. In this Review we present a holistic and interdisciplinary examination of V2G services, highlighting the wide range of potential benefits as well as the challenges slowing the technology's evolution from niche trials to mainstream adoption. We find that benefits tend to be siloed by value proposition and stakeholder while the challenges tend to stem from stacking multiple values and connecting multiple stakeholders.
Consequently, we identify key areas for future research, industry and policy activities that will accelerate and smoothen the realisation of V2G’s potential as an essential pillar of clean transport-electricity systems.

\end{abstract}
\begin{document}

\flushbottom
\maketitle

\thispagestyle{empty}

\section*{Highlights}

\begin{itemize}
    \item Transport and electricity impacts of V2G are reviewed from social, economic, and technical perspectives.
    \item Economically, V2G leverages existing, underutilised assets.
    \item Technically, V2G provides the flexibility required for the transition to a distributed and decarbonized electricity system.
    \item Social factors of status and self-reliance materially affect decisions to engage with V2G services.
    \item Uncertainty, in financial returns and vehicle availability, is the current main barrier.
    \item Immature relationships across the transport and electricity sectors are a systemic challenge to mainstream adoption.
\end{itemize}

\section*{Keywords}
mobility, transportation, distributed energy resources, electric mobility, vehicle-to-grid, vehicle grid integration

\vspace{2mm}
{\bf Word count:} 5448

\section*{List of abbreviations and units}

\begin{itemize}
    \item AC - Alternating Current
    \item DC - Direct Current
    \item CCS - Combined Charging System 
    \item CHAdeMO - CHArge de MOve
    \item DER - Distributed Energy Resources
    \item EV - Electric Vehicle
    \item ICE - Internal Combustion Engine
    \item ICT - Information and Communications Technology
    \item SAE - Society of Automotive Engineers
    \item USD - United States Dollar
    \item V0G - Behaviour controlled charging
    \item V1G - Managed charging
    \item V2B - Vehicle-to-Building
    \item V2G - Vehicle-to-Grid
    \item V2H - Vehicle-to-Home
    \item V2X - Vehicle-to-Anything
    \item VGI - Vehicle Grid Integration
\end{itemize}

\section{Introduction}

Access to electricity and far-ranging mobility are defining features of modern life, especially in wealthy nations. Historically, these services have been provided by largely independent systems and sectors but the accelerating uptake of electric vehicles (EVs) driven by a wide range of factors including compatibility with living arrangements and life stage \cite{li_review_2017}, superior driving performance \cite{xu_moving_2020}, cost minimisation and environmental concern \cite{graham-rowe_mainstream_2012} is creating systemic interdependencies between the electricity and automotive transport sectors towards an ultimate aim of decarbonisation and increased flexibility of energy use, referred to as ``sector coupling'' \cite{robinius_linking_2017}.
EVs clearly depend on the electricity system to fuel their motion. Studies focusing on the United States have shown that EVs may increase electricity demand by around one quarter and that a 25\% penetration of EVs could increase peak demand by 19\% \cite{thompson_perez_2020}. EVs however also present a significant potential upside for the electricity system in that their batteries could be a valuable source of flexible, fast access energy storage necessary for increased renewable energy penetration.
The effects of sector coupling are not limited to the integration of technical systems, but involve a profound reshaping of people’s conceptions of these essential services, as well as of the techno-economic systems governments and businesses use to provide them. 

\begin{figure}[!ht]
\centering
\includegraphics[width=0.8\linewidth]{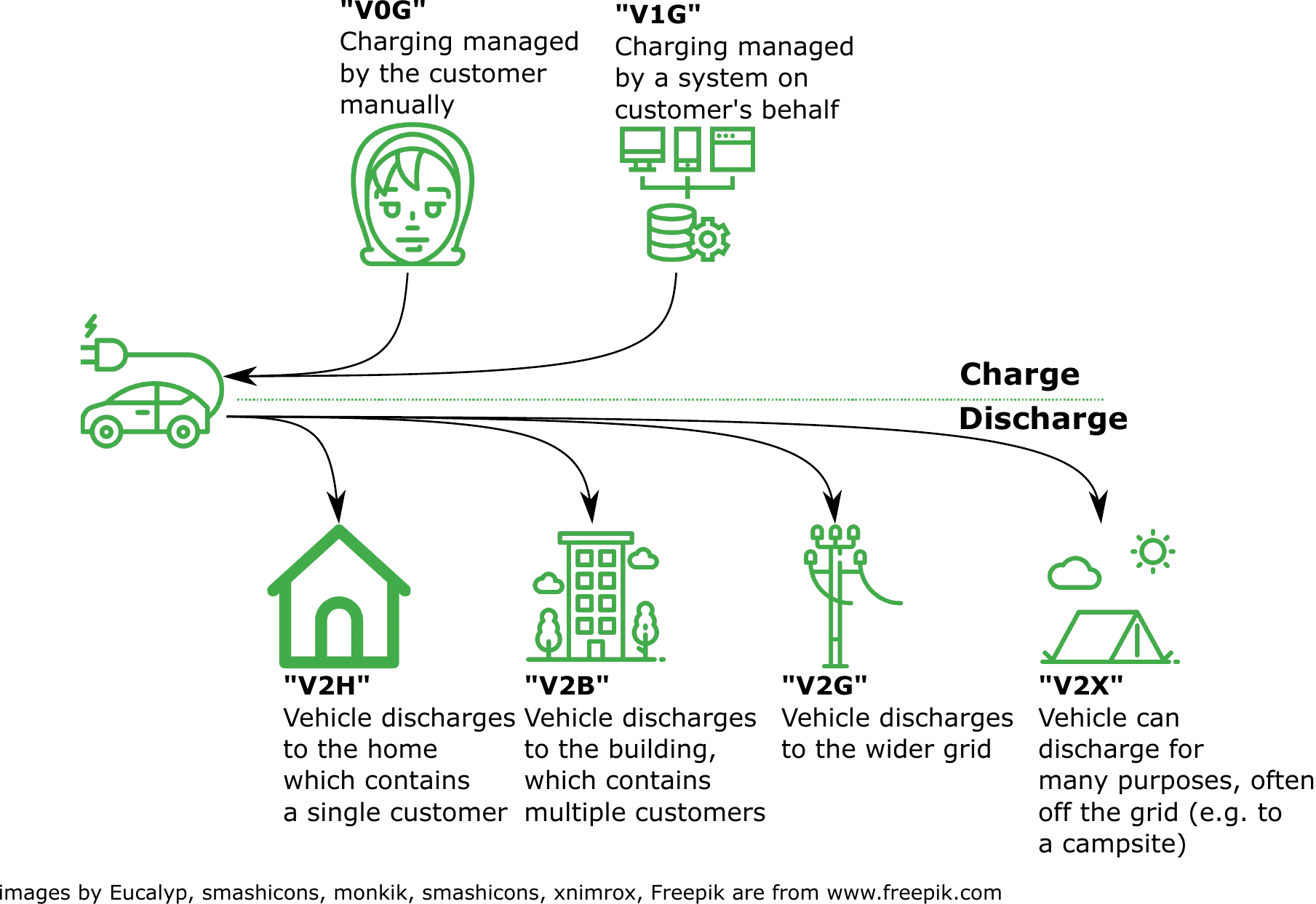}
\caption{Methods for managing EV charging and discharging. For charging, options are: behavioural control (V0G), where customers manually modulate charging; and managed charging (V1G or G2V), where chargers have  algorithmic control. For discharging, configurations include: Vehicle-to-Home (V2H), where the vehicle's energy is consumed within the home; Vehicle-to-Building (V2B) where the energy is consumed in a single building (possibly consisting of several homes); Vehicle-to-Grid (V2G), where the energy provides services to the wider electricity grid; and Vehicle-to-Anything (V2X), specifies wider uses of discharge capacity (for example for camping). In this Review we use V2B to cover V2H and V2B.
}
\label{fig:VGI}
\end{figure}

The act that couples the electricity and transport systems - physically, financially, and experientially - is power flowing between the electricity grid and vehicles. Typically, power flows from grid to vehicle to charge the vehicle for driving. The large power and energy capacities of EVs means that it will become vitally important to manage this charging to reduce the potential negative impacts on the grid. As shown in Figure~\ref{fig:VGI}, this can be achieved by influencing the manual charging behaviour of customers (referred to as V0G) or through automated ICT systems that respond to user preferences and control and/or price signals (referred to as V1G). These configurations are limited to unidirectional charging, whereas  here we focus on an extended configuration that manages both the charging and discharging of EV batteries. As shown in Figure~\ref{fig:VGI}, EVs can discharge power into a number of systems, ranging from national electricity grids through to buildings to appliances. Unless otherwise specified, we use the term Vehicle-to-Grid (V2G) to refer to all these applications, focusing on the key distinction that only V2G-enabled vehicles can, when plugged in to a V2G-enabled charger, discharge power from their battery for external uses \cite{noel_vehicle--grid_2019,tuttle_evolution_2012}.

Conceptually, the core proposition of V2G is simple: EVs have large, underutilised batteries that should be enabled to discharge power to provide a range of services beyond just powering the drive train. This is particularly pertinent because many vehicles spend considerably longer parked and connected to chargers than is required for recharging their battery.
The V2G concept has been demonstrated in laboratories since the 1990s \cite{kempton_electric_1997} and trialled in real-world settings \cite{everoze_v2g_2018}, but widespread deployment of V2G services has yet to follow. 

\begin{figure}[!ht]
\centering
\includegraphics[width=0.7\linewidth]{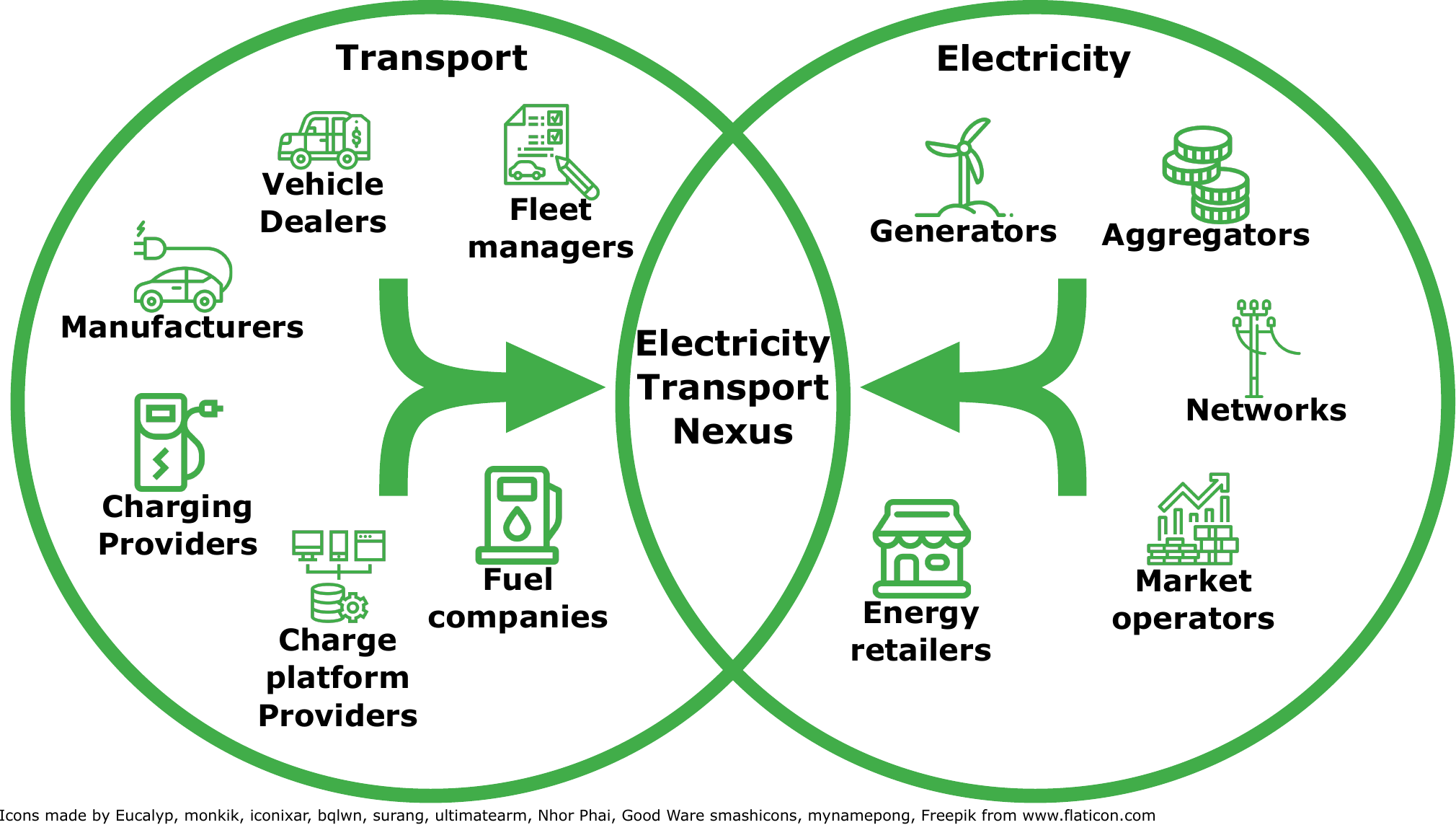}
\caption{The transportation-electricity nexus. Transport sector stakeholders include: vehicle manufacturers, who provide V2G capable vehicles and warranties; charging providers, who manufacture bi-directional EV charging hardware; charge platform providers, who provide V2G management software; vehicle dealers, who may be the first to communicate V2G services; and fuel and fleet management companies, who may see V2G as an opportunity or barrier. Electricity sector stakeholders include: generators, who provide energy and ancillary services and may be a competitor or customer of V2G services; electricity networks, who connect V2G chargers and may require V2G services; market operators, who manage the power system and market and define market rules; retailers, who manage supply of energy on behalf of customers and may also manage V2G chargers; aggregators, who aggregate large numbers of small flexible generation and loads such as V2G into markets, and; electrical contractors, who play a critical, yet oft overlooked, role in rolling out new devices into the field.
}
\label{fig:nexus}
\end{figure}

We posit that the failure of V2G to evolve beyond the niche stems from the fragmentation of knowledge and experiences into traditional disciplinary silos and across transport and electricity systems stakeholders (Fig.~\ref{fig:nexus}), each of which have their own practices, technologies, products, skills, procedures, established user needs, regulatory requirements, institutions and infrastructures \cite{hoogma_experimenting_2002}.
A review of 50 V2G trials from around the world reported that while essentially all (97\%) considered technical performance less than half (40\%) considered commercial issues and less than a third (27\%) considered social aspects \cite{everoze_v2g_2018}.
Similarly, a review of V2G publications between 2015 and 2017 found that research was heavily weighted towards science and engineering with few contributions from social science, economics, or arts and humanities \cite{sovacool_neglected_2018}. 


\begin{figure}[ht]
\centering
\includegraphics[width=0.7\linewidth]{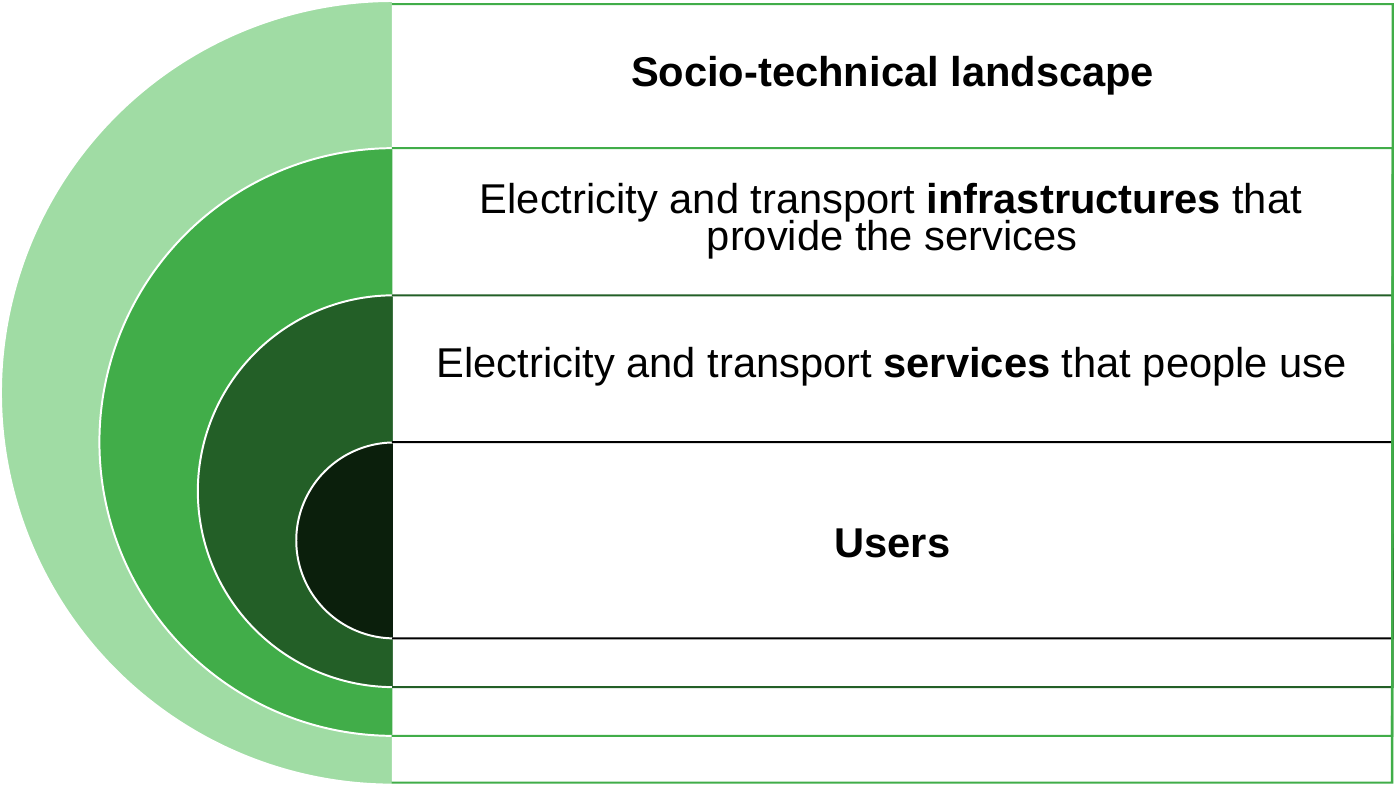}
\caption{Our framework for analysing V2G. Users have certain needs and wants, which are serviced by the electricity and transport system. These systems rely on physical and metaphysical infrastructures in order to function, and both users and the systems are embedded in a broader socio-technical landscape that contains other people and systems.
}
\label{fig:framework}
\end{figure}

Our Review expands on previous reviews via its scope, taking a cross-disciplinary view. We highlight benefits of V2G demonstrated in narrowly targeted research and trials as well as the challenges facing widespread deployment from interdisciplinary and/or cross-sector interdependencies. In order to cover this breadth of content, our Review is informed by not only the academic literature across many disciplines but also industry reports from trials, regulatory and government policies, and dozens of frank, off-the-record conversations with stakeholders.

To make sense of the broad and varied nexus of the transport and electricity systems we developed the framework depicted in Figure~\ref{fig:framework}. Our framework places end-users of transport and electricity services at the centre. The services and value they provide to the people and organisations using them are explicitly called out in the next layer. The value of V2G services extends beyond the financial realm to include independence, convenience, status, affordability, environmental concerns, community-mindedness, and trust in institutions and service providers. The next layer in our framework considers the infrastructures involved in providing services. These include not only physical network infrastructures and their hardware components, but also the markets, transactions, rules, conventions, software and standards. Lastly, these infrastructures, services and users all exist within a broad socio-technical landscape characterised by uncertainty and change. Our review therefore considers the landscape pressures wrought by broader issues including the energy transition and climate change.

The rest of this Review is structured as follows: we begin by presenting the emerging and established benefits of V2G, we then discuss the challenges that are impeding the growth of V2G, before concluding by forming a holistic assessment to provide our outlook of how V2G can progress from niche to normality, and specifying some of the tasks that are likely to be needed on this journey.

\section{Benefits}

V2G offers a remarkable range of potential benefits \cite{noel_vehicle--grid_2019,noel_willingness_2019}. In this Review we categorise both these benefits and the challenges by recipient: vehicle users, the electricity system, and society as a whole.


\subsection{For vehicle users}

Of the wide range of benefits that V2G may deliver to vehicle owners, three stand out: financial, energy resilience, and status.
The financial benefits stem from better energy management, which can involve three aspects \cite{pearre_review_2019}. Firstly, many owners of EVs own rooftop solar systems \cite{cohen_q-complementarity_2019} and moreover rooftop solar ownership can be locally very high \cite{coffman_integrating_2017}; subsequently, people can use their vehicles to store excess daytime solar energy for later use thereby reducing their reliance on grid electricity and increasing the return on their solar investment \cite{noel_vehicle--grid_2019,chen_strategic_2020,kam_smart_2015,sommerfeld_residential_2017}. Such an arrangement is central to the value proposition of V2B and V2H. Secondly, V2G can generate income for vehicle owners by providing services to the grid, such as supporting greater integration of renewable energy, energy price arbitrage, frequency control, or congestion management \cite{pearre_review_2019,mwasilu_electric_2014,hernandez_primary_2018,rezkalla_comparison_2018,zecchino_large-scale_2019,kaur_coordinated_2018,marinelli_validating_2016,izadkhast_design_2017,moghadam_distributed_2015,esther_dudek_electric_2019,mcdougall_ev360_0000}. Thirdly, capital-efficient use of EV batteries can compete favourably with existing fossil-fuel generators \cite{thompson_perez_2020} which should place downward pressure on wholesale electricity prices and translate to lower costs for all electricity users, including owners of V2G EVs. 

The second major benefit of V2G for vehicle owners is that the energy stored in their vehicles can be used for backup power supply \cite{kwon_high_2015,steward_critical_2017}. Studies have shown that people place a high value on having backup power, even if it comes at the expense of trading opportunities \cite{thiebaux_consort_2019}. Bolstering energy resilience is a growing concern as climate change increases the quantity and severity of natural disasters, and it is therefore understandable that trials have been focused on regions that are particularly vulnerable to disasters such as Japan. These trials have shown that an EV can power an average home for four to 12 days, particularly when combined with rooftop PV generation \cite{tuttle_plug-vehicle_2013,thompson_perez_2020} and that, on a larger scale, vehicle power can be integrated into microgrids \cite{noel_willingness_2019,thompson_perez_2020} at community facilities such as hospitals, community halls, and convenience stores, which provide essential services and in some cases life support \cite{noel_beyond_2018,thompson_perez_2020}. The mobile nature of EVs enhances their utility for backup as energy can be transferred from chargers to locations without electricity supply.

The third and to date least understood benefit of V2G to users is in affirming self-identity through symbolic meaning. Status-seeking is well established as a form of conspicuous consumption \cite{bronner_conspicuous_2018} in car purchasing decisions and V2G may speak to both nature-oriented or wealth-oriented people \cite{perlaviciute_influence_2015} to signal their environmental awareness or technology savviness \cite{noel_vehicle--grid_2019,noel_willingness_2019,noel_conspicuous_2019}. This role of symbolism is reflected in studies that have found customers to have a strong preference for charging their EVs with renewable energy, with some customers willing to pay an additional USD0.61 per hour for renewable energy at public charging stations \cite{nienhueser_economic_2016}, and that customers with solar power had an increased preference for charging at home \cite{jabeen_electric_2013}. The issue of status is not yet settled, and deserves further research, as other studies have found that political orientation plays a significant role in peoples’ willingness to pay for EVs \cite{arpan_politics_2018} as well as their concerns about the environmental and social impact of lithium-ion batteries \cite{berkeley_assessing_2017,parsons_willingness_2014,noel_fear_2019}. Others have found that V2G could be perceived to be in conflict with traditional meanings of the car such as freedom \cite{noel_fear_2019,kline_users_1996,graham-rowe_mainstream_2012,steg_car_2005}. The weight of this consideration was emphasised by the finding of one study that people may be happy to participate in V2G even without revenue, if the design of the scheme did not impair their sense of freedom \cite{geske_willing_2018}. The significance of freedom to decision-making could present an opportunity to turn conversations to the electricity independence provided by V2G. This could favour VGI configurations more weighted towards home energy management than providing grid services.

\subsection{For the electricity system}

The technical benefits of V2G for the electricity system have to date been the primary focus of research \cite{sovacool_neglected_2018,everoze_v2g_2018}. The studied benefits fall into two broad categories: helping balance power supply and demand, and helping manage the operating state of distribution networks.
The application of EVs to these challenges has much in common with the use of stationary batteries, and both are significant components in the electricity system's transition to a more decentralised, digitised, and decarbonised electricity system \cite{thompson_perez_2020}. Their similarities enables opportunities for leverage, as well as situations of competitive tension.

Balancing power supply and demand is a critical element of maintaining a secure electricity system \cite{alexandra_von_meier_electric_2006}. The strategies for achieving this can be divided into two classes: the first is focused on maintaining the balance over the time horizon of seconds, while the second is concerned with the balance across time periods of hours to days (intra-day). V2G has been demonstrated to assist with both applications.

\begin{figure}[ht]
\centering
\includegraphics[width=0.6\linewidth]{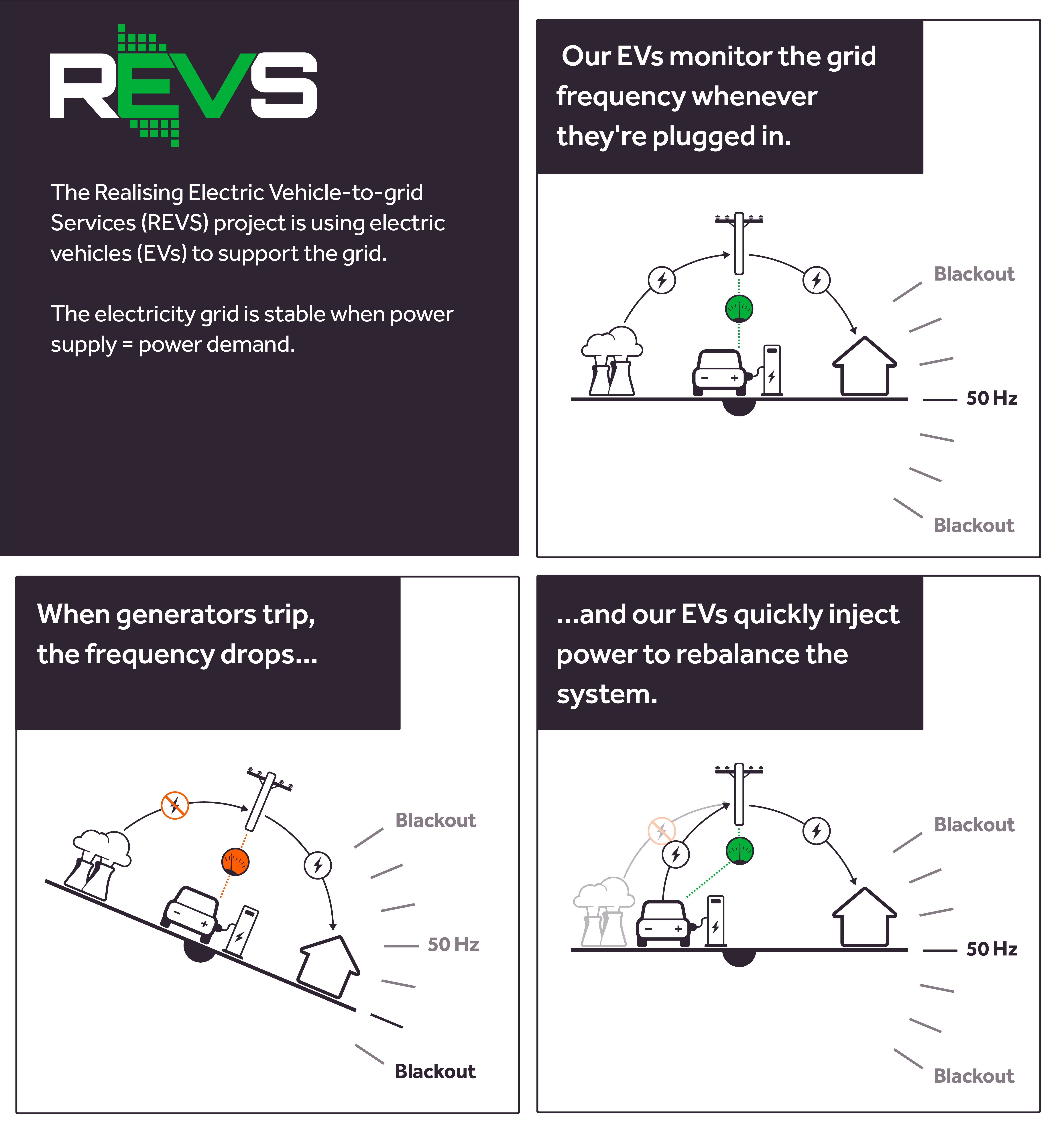}
\caption{Frequency control application of V2G, maintaining the balance of power supply and demand on a sub-second basis.
}
\label{fig:frequency}
\end{figure}

The standard way of monitoring sub-second imbalances in supply and demand is by measuring the frequency of the voltage waveform, which is directly related to the power balance of synchronous generators \cite{alexandra_von_meier_electric_2006}. While power electronics-coupled devices such as batteries do not have a physical response to variations in frequency, their chemistry and electronic control systems allow them to execute programmed responses to variations within hundreds of milliseconds \cite{kempton_electric_1997}. 

Frequency control is in turn typically delivered in two parts: small mundane variations are managed continuously through regulation responses, while major emergency imbalances, such as an unexpected disconnection of a large generator, are managed with contingency responses.
Regulation control is typically implemented using a centralised control signal transmitted by system operators. Numerous trials have integrated V2G EVs with these systems \cite{hernandez_primary_2018,rezkalla_comparison_2018,zecchino_large-scale_2019,kaur_coordinated_2018,marinelli_validating_2016,izadkhast_design_2017,moghadam_distributed_2015,hashemi_frequency_2018,black_angeles_2018}. A systemic challenge with this control architecture is the potential for time delays due to communications and processing lags. These have been found to be on the order of 3-6 seconds, which are acceptable for frequency regulation services \cite{kempton_test_2008,hashemi_frequency_2018,black_angeles_2018}.

V2G has also been applied to the provision of contingency frequency control \cite{izadkhast_aggregate_2014,liu_enabling_2018} (illustrated in Figure~\ref{fig:frequency}) for which fast timing requires local control. 
Contingency services have a striking advantage for EVs in that vehicles are paid for being available (plugged in) but are rarely called upon to (dis)charge because contingencies occur very rarely. These services thereby create minimal impositions on batteries or users \cite{kempton_test_2008,black_angeles_2018}.
Local frequency control has been implemented using algorithms based on absolute measurements of frequency (droop control) as well as algorithms based on rate of change of frequency (synthetic inertia) \cite{rezkalla_comparison_2018}.
Rezkalla et al. compared these control approaches through simulations and experiments and found that while inertia-based control may provide better performance, it is more complex and challenging to implement in the real world, while droop-based control provides acceptable performance with simpler implementation \cite{rezkalla_comparison_2018}.

EVs may also have a major role to play in the intra-day balancing of supply and demand. Because EVs are a net load on the electricity system (with energy expended on transportation) their primary contribution comes from modulating the timing of their charging. This contribution may be very significant due to the high power and energy demands of EVs coupled with the high degree of flexibility of when this charging occurs \cite{esther_dudek_electric_2019}.
Trials have demonstrated that this flexibility can be utilised even through the simple means of sending customers EV specific price signals which they can then choose to respond to (V0G) \cite{mcdougall_ev360_0000}. Such rudimentary approaches run the risk of concentrating EV charging to such a degree that it creates new peak loads. The EV360 trial showed 99\% compliance with their price signal and highly co-incident charging \cite{mcdougall_ev360_0000}. This impact is much the same as how co-incident generation can cause market prices to drop at times of surplus \cite{hirth_market_2013}. 
Automation of pricing signals, through `smart' internet connected chargers (V1G) can facilitate greater flexibility while avoiding co-incident charging. Numerous studies have used such approaches to align vehicle charging with renewable energy generation, both from rooftop solar systems and utility renewable generation \cite{pearre_review_2019,shi_integration_2020,mwasilu_electric_2014,quddus_modeling_2019,hernandez_primary_2018,rezkalla_comparison_2018,international_renewable_energy_agency_innovation_2019,szinai_reduced_2020,thompson_perez_2020}. This is not only beneficial for the electricity system but builds on the appeal of EVs to environmentally conscious consumers.
V2G extends the possibilities for intra-day balancing, particularly by using EV batteries to store excess daytime solar and power homes during evenings. This application has received less academic research but is more intuitive to users and is therefore prioritised in vehicle manufactures' story telling.

In addition to system-wide balancing benefits, V2G offers significant benefits for managing the operating state of distribution networks. Specifically, V2G EVs can adjust power flows on local networks such that they respect the physical and operational voltage and thermal limits of all network elements, deliver high power quality with low harmonics and near-unity power factors, and maintain the system strength of the network. 

Forward-looking scenario modelling has made it clear that V1G charging orchestration will be fundamental in accommodating mass EV charging within the limits of distribution networks \cite{dubey_electric_2015,yilmaz_review_2012} as unmanaged charging could increase peak demand by 50-100\% \cite{7051704}. 
V2G enables EVs to contribute to `peak shaving' by discharging power at peak times \cite{telaretti_battery_2016,li_cost-benefit_2020,uddin_review_2018}. Field trials of V2G have demonstrated peak shaving of up to 30\% \cite{li_cost-benefit_2020,gadh_demonstrating_2018} and utilities are now incorporating V2G into demand response markets \cite{california_iso_energy_2020}. 

V2G can also improve power quality by providing reactive power support to manage voltage and improve power factors \cite{li_mpc_2019,brinkel_impact_2020,wang_two-level_2019,dong_charging_2018,alam_effective_2015,traube_mitigation_2012,clement-nyns_impact_2009}. These services are delivered by the power electronic converters within vehicles or chargers, without draining the battery and, when provided by chargers, without a dependence on vehicle availability. These services are naturally combined with broader decentralised energy integration strategies to reduce distribution losses \cite{restrepo_three-stage_2016,alam_effective_2015,lopes_integration_2010,knezovic_phase-wise_2016}. 

Finally, V2G can bolster system strength, being the power system’s tendency to remain within safe operating conditions even under stressed conditions and to recover adequately after disturbances. Studies have demonstrated V2G chargers acting as virtual synchronous machines (providing synthetic inertia and frequency control that enables seamless transition to islanded operation during outages on the grid) \cite{suul_virtual_2016} as well as providing fault current contributions and post-fault reactive power support \cite{islam_short-term_2020,katic_impact_2019}. These are in addition to the broader opportunities for V1G to contribute to lifting minimum demand, whose trajectory in certain regions towards zero is creating unprecedented control challenges for system operators, including disabling protection equipment \cite{australian_energy_market_operator_minimum_2020}.

\subsection{For society}

V2G also promises significant public benefits arising from its role in decarbonising both the transport and electricity systems. The core argument of V2G is the utilisation of existing battery assets residing unused in parked EVs to provide services that are needed for the operation of an electricity grids transition, as we have detailed in the preceding sections. This efficient use of assets ought to reduce the costs of operating the electricity system and flow on to lower prices for end-users. Lower prices particularly benefit low income households, which generally spend a disproportionately high percentage of their income on electricity and gas \cite{australian_council_of_social_service_energy_2018}. 

Mainstream adoption of V2G should also catalyse significant health and climate co-benefits by accelerating the replacement of internal combustion engine (ICE) vehicle stock with EVs and in doing so reduce greenhouse gas emissions and air pollution \cite{noel_vehicle--grid_2019,parsons_willingness_2014,von_stackelberg_public_2013}. This is a significant benefit as in the US alone, climate change and air pollution attributable to electricity generation and transport cost an estimated USD470 billion per year \cite{noel_vehicle--grid_2019}. In Australia, there are an estimated 684 deaths per year as a result of transport emissions, costing USD6.4 billion per year using a local estimate of USD10.4 million of the value of a statistical life \cite{anenberg_global_2019}. EV adoption is also a necessary step towards autonomous vehicles which are claimed to be able to reduce congestion and road accidents, improve fuel efficiency and make mobility more accessible \cite{bahamonde-birke_systemic_2018}. A further societal benefit of a transition away from fossil transport fuels is reduced dependence on oil \cite{berkeley_assessing_2017}.

The adoption of EVs and the accompanying integration of charging into electricity grids will need a well-planned approach that takes advantage of opportunities provided by EVs \cite{mwasilu_electric_2014}. From this point of view, strong adoption of EVs could make V2G and other forms of VGI necessary for avoiding unmanaged growth of peak demand and other ill effects, rather than being niches with uncertain futures. This emphasises V2G's position as both an enabler of accelerated EV adoption, and a solution to the impacts on the electricity grid of more EVs.


\section{Challenges}

In contrast to the clearly identifiable benefactors of V2G outlined above, the challenges of moving V2G from niche into the mainstream tend to arise from the interaction of numerous systems and stakeholders in yet to be crystallised configurations. It requires a series of technical, economic and social relationships - spanning the electricity and transport sectors - that may not currently exist.

\subsection{For vehicle users}

There are at least five major challenges that directly affect the value proposition of V2G for vehicle users: costs, battery degradation, awareness, control, and charger efficiency. 

Currently, the major cost barrier to V2G arises from the cost of V2G-capable chargers \cite{thompson_perez_2020}. This cost is due to additional hardware and software complexities \cite{yilmaz_review_2012,nelder_reducing_2019}, their current low volume of sales, and significant `soft costs' arising from installations, including permitting delays, utility connection requests, dealing with fragmented regulations and redesigns owing to regulatory and industry learning curves \cite{nelder_reducing_2019}.
This situation is exacerbated by CHAdeMO DC charging being the only standard that explicitly supports V2G, further limiting uptake and increasing costs due to the smaller market. Costs are expected to decrease along trajectories similar to batteries and (non-V2G) EVs \cite{noel_vehicle--grid_2019,berkeley_assessing_2017} as production ramps up, the Combined Charging System (CCS) standard is updated to become the second standard to facilitate V2G \cite{charin_ev_grid_2020}, and AC options become available (such as the updated SAE J3068 standard in North America). 

Exacerbating the cost barrier, there are significant uncertainties in the financial rewards available to customers through V2G services. One study in the United States showed that the difference between consumption tariffs and frequency control prices, as well as taxation, made the provision of frequency services unprofitable under current rules \cite{kempton_test_2008} while a Danish study concluded that frequency control services at best broke even, even when tax implications were excluded \cite{christensen_project_2018}.

A concern for proponents of V2G that we argue is more urgent relates to the perceived impact on EV battery lifespans that could arise from their greater utilisation. 
Research on battery degradation suggests that numerous factors have more pronounced impacts on battery degradation than the number of cycles, including: both high \cite{electric_vehicle_wiki_battery_0000,ouyang_impact_2020} and low \cite{schimpe_comprehensive_2018,omar_lithium_2014} temperatures, with one study finding battery life halving due to an increase of 10$^{\rm o}$C \cite{stroe_operation_2016}; high current draw rates \cite{ouyang_impact_2020}; time spent at the extremes of state of charge \cite{stroe_operation_2016,schimpe_comprehensive_2018,omar_lithium_2014}, with even a moderate reduction in cycle depth from 90\% to 70\% increasing battery life by 8-10 months in one study \cite{stroe_operation_2016,schimpe_comprehensive_2018} and approximately 1.8 years in another \cite{hoke_maximizing_2013}. 
Taking all of these factors into account, `smart' charging strategies, including V1G and V2G, have been shown to be able to significantly reduce battery degradation, more than compensating for the extra cycling loads \cite{hoke_maximizing_2013,tan_coordinated_2017,uddin_possibility_2017,smith_comparison_2012}.
While these impacts should be quantified further, our key point is that EV owners' are very sensitive to this issue \cite{geske_willing_2018,noel_willingness_2019} and that their perceptions of this risk are what V2G proponents must address in their communications.

The role of communication is also central to tackling the challenge of general unfamiliarity and ambivalence that the public have towards V2G \cite{noel_vehicle--grid_2019}. Trials such as the Parker project found that ``a proper introduction to V2G'' is required to ``build acceptance'' \cite{andersen_parker_2019}. Low awareness may also present an upside opportunity, in that people lack preconceptions against V2G \cite{geske_willing_2018}. What is clear, is that raising awareness and understanding is absolutely critical among current and potential users because in the end ``It is their EV, their battery, and their power that is being used, sold, stored, and served'' \cite{noel_vehicle--grid_2019}. 

Proponents of V2G must keep in mind that V2G is, at least for the foreseeable future, a side issue in people buying EVs. The vehicle's primary purpose is providing flexibility and convenience for owners \cite{noel_willingness_2019,berkeley_assessing_2017}. The potential – perceived or real – of V2G to negatively impact these values by constraining usage and draining batteries can create strong rejection of V2G participation \cite{noel_willingness_2019,andersen_parker_2019}, particularly as these issues focus buyer's attention on issues of range anxiety and recharge times \cite{berkeley_assessing_2017,franke_interacting_2013,pearre_electric_2011}.
These concerns mirror research on stationary batteries \cite{ransan-cooper_frustration_2020} in identifying the ceding of control to orchestration engines as a challenge to both flexibility and cyber-security \cite{bailey_anticipating_2015,geske_willing_2018,noel_vehicle--grid_2019,noel_willingness_2019}. A particular research finding to note is that drivers see high inconvenience costs to signing V2G-EV contracts \cite{parsons_willingness_2014}, even when contract terms are easy to fulfil \cite{noel_vehicle--grid_2019}. Fortunately, it may be that the way the contracts are framed are the concern, rather than V2G participation itself \cite{noel_willingness_2019}. A study of more than 3,000 people in the United States recommended contracts either adopt earn-as-you-go arrangements, without a fixed term, or advanced cash payments in return for signing onto a contract term. These are preferred because people tend to associate a high inconvenience cost to V2G participation and associate a high uncertainty with earning money from V2G \cite{parsons_willingness_2014}. In another study people applied a heavy discount of 53.5\% to projected revenue, which is high even by the standards of energy efficiency \cite{hausman_individual_1979}. Some commercial V2G providers are starting to address this with day-ahead scheduling of EV use, contracts that oblige customers to participate in only a proportion of demand response events, and by providing an override option.

Lastly, the introduction of additional electrical components unavoidably introduces energy losses. Losses in bidirectional chargers are of direct and particular relevance to V2G, as energy lost in charge/discharge cycles are paid for by EV owners without providing any benefits. This is currently a poorly understood issue \cite{noel_vehicle--grid_2019,shirazi_comments_2018,apostolaki-iosifidou_measurement_2017}. One study found that losses occurred predominantly in the power electronics used for AC/DC conversion, and that efficiency depended on the charging rate, battery state of charge, and whether it was charging or discharging, with differences in efficiency amounting to at least 7\% \cite{apostolaki-iosifidou_measurement_2017}. This means that just as V2G charger algorithms can be configured to protect battery health, they could also optimise efficiency.

\subsection{For the electricity system}

A challenge for integrating V2G (and EVs generally) into the electricity system is their variable availability at charge points. This is pronounced with V2G because many of the aforementioned services are critical to the system security and because variable availability is the key differentiator between EVs and stationary batteries, which are often vying to service the same shallow frequency and network support markets.
This places the burden onto V2G solutions to reduce the risks of availability to such an extent that the efficiency of leveraging an existing mobility asset for electricity system services outweighs the simplicity of procuring a dedicated stationary battery.

Aggregation of many EVs can go some way towards mitigating these risks \cite{izadkhast_aggregate_2014}.
Doing so in practice is a complex task that is usually undertaken by a dedicated entity, called an aggregator \cite{burger_value_2016,international_renewable_energy_agency_innovation_2019-1,best_understanding_2019}. Aggregators provide a range of functions in the energy system including grouping diverse agents together to take advantage of economies of scale and scope and managing risks \cite{burger_value_2016,international_renewable_energy_agency_innovation_2019-1}. Aggregators face technical, commercial, and regulatory barriers to their success  \cite{ran_maximizing_2019,international_renewable_energy_agency_innovation_2019-1}.
Technically, aggregators must present diverse resources as a single aggregated sum to their partners. This requires multiple values to be co-optimised \cite{tong_energy_2017}, and relies on granular, accurate forecasts \cite{ran_maximizing_2019,international_renewable_energy_agency_innovation_2019-1}. This may be especially challenging when there are few resources, for example as may be seen in some distribution network congestion scenarios \cite{ran_maximizing_2019}. Value stacking can be similarly complex. Each value stream may have a different target, such as availability, energy or power, and may require different actions at the same time in order to realise them \cite{tong_energy_2017}.
Furthermore, some prospective value streams, such as resilience and voltage management, require market reform in order to become accurately valued.
Commercially, aggregators must navigate a complex landscape. They must present diverse resources to markets as a single risk-managed product while managing their customer's diverse needs \cite{burger_value_2016}. They also face significant startup costs, ICT systems and minimum participation thresholds \cite{burger_value_2016,international_renewable_energy_agency_innovation_2019-1}. For inexpert new-entrant agggregators this landscape can be especially challenging to navigate \cite{burger_value_2016}. 
Participation requirements can be a large barrier to aggregators. Standards may place high requirements on each individual participant in market value streams, increasing the cost of aggregation \cite{burger_value_2016}. Uptake of aggregation is improved through opening up market structures, reducing complexity, and lowering barriers to participation \cite{international_renewable_energy_agency_innovation_2019-1}.  

Lastly, a technical area requiring further development is improving the harmonic distortion of the power electronic converters in chargers, which currently can be as high as 31\% \cite{bass_impacts_2013} and can cause increased heating and reduced life expectancy of equipment such as transformers \cite{taylor_evaluations_2010}.

\subsection{For society}

V2G is likely to intersect — and interact — with a range of socio-political barriers if it is able to succeed in moving from its current protected niche to become a normal part of the wider electricity and automotive transport `regimes’ or systems. Its success is not simply a case of technological advancement, nor enough EV owners opting in; it also requires disruption of an existing regime and its underlying policy, politics and power \cite{geels_regime_2014}. Disrupted incumbents include fossil fuel companies, internal combustion engine vehicle manufacturers, dealerships, mechanics, refuelling stations, traditional energy companies, as well as government fuel tax revenue, some or all of which may generate resistance \cite{berkeley_assessing_2017,noel_vehicle--grid_2019}. 

There is also a risk that V2G contributes to perpetuating conventional (private) automobility \cite{noel_vehicle--grid_2019}, which causes congestion that reduces access to services, employment and social support \cite{noel_vehicle--grid_2019} as well as causing death and injury in accidents and physical inactivity leading to obesity, diabetes and heart disease \cite{woodcock_energy_2007}. Automotive transport also contributes to land use for roads and parking, noise, energy dependence and the undermining of communities \cite{hoogma_experimenting_2002}, which disproportionately impacts low income, vulnerable and non-white communities \cite{VERBEEK2019100340}. Other interventions, including active travel, public transport, and `last mile' options will be needed to address these challenges and ``extend the range of human mobility'' \cite{ulrich_estimating_2005}. On the other hand, as noted in the benefits section, autonomous vehicles are touted as solutions, and EVs are a necessary step in that direction. Examination of these society-level issues through lenses of long term urban and transport planning and public health is a good-practice approach, but has the potential to obscure or take for granted the needs of the clean energy transition. From this point of view, the focus on the electricity and automotive transport nexus in this Review is one perspective, but a broader view that implicates urbanisation, climate change, public health and other big picture concerns reveals even more complexity.

Further to the problem of the clean energy transition, policies that promote EVs and related technologies have the potential to generate criticism as EV ownership appears to skew towards people on higher incomes \cite{sovacool_income_2019} and men \cite{sovacool_are_2019}. For social acceptance and society-wide benefits to flourish it has been suggested that technologies like V2G should be developed and deployed through a co-design process that incorporates citizens' aspirations for the future and provides a way for these to be implemented with transparency and fairness \cite{dahlgren_digital_2020,ransan-cooper_frustration_2020}.

\subsection{Conclusions and outlook}

EVs create a strong coupling between the electricity and automotive transport sectors. V2G enhances this by allowing the vehicle battery's to be used for external uses when the vehicle is parked and plugged into a suitable charger. There are many potential benefits to V2G - accruing to EV owners, the energy system, and society as a whole.
EV owners' benefit streams include financial, self-reliance, resilience, and status. The energy system benefits due to the flexibility of V2G to respond to system needs using an existing asset. This flexibility can manage distribution network congestion, integrate renewable generation, manage market price, and assist controlling frequency. To society, V2G provides the means to use EVs to reduce emissions and enhance resilience in the electricity as well as transport sectors.

Tempering these benefits are challenges and risks, which tend to arise from the stacking of multiple values and connecting multiple stakeholders. Primarily, potential users are as yet unaware of V2G and its potential benefits. V2G is costly and not currently well supported by charge standards. Also, V2G may reduce the convenience and freedom offered by EVs to their users, or at least have the perception of doing so. The additional energy usage of V2G has the potential to negatively impact vehicle battery health, especially if not managed well. Energy system participation also faces challenges. Participation may require large numbers of participants and costly equipment in each installation. Particularly in the distribution network, participation may require accurate forecasts of small numbers of EVs. These barriers are challenging for new entrant service providers to navigate. 

Niches are spaces that protect innovations from outside pressures while their viability is being demonstrated through learning, using and interacting \cite{rip_technological_1998}. The eventual goal, however, is for the innovation to move from being a discrete technical artefact to being linked into a socio-technical regime of industry and user practices \cite{rip_technological_1998}. In the case of VGI, this has occurred to a modest extent in Japan where, since the T\={o}hoku earthquake, energy supply shortages, and consequent government actions, V2B has been used to provide critical backup power \cite{kobashi_potential_2020}. In other countries a market-led approach is envisaged, assisted by increases in the adoption of EVs, digitisation, decentralised energy prosumers, and variable renewable energy, giving rise to new challenges and opportunities. These trends are inexorably linked to the futures of both electricity and mobility, which are characterised by uncertainty. These types of sustainability transitions are particularly challenging due to contested normative judgements on what to do or what technology to back, and split incentives between local and global, individual and collective \cite{geels_ontologies_2010}. As a result, reliance on the market might not be sufficient for the normalisation of technologies like V2G: public authorities, social movements and public opinion are very important in directing and mobilising V2G out of its niche \cite{geels_ontologies_2010} and towards a configuration that works best for EV owners and other users - not necessarily mirroring what is envisaged by electricity sector actors.

The question then is how niche activities and broader policies could best position V2G to successfully evolve into the mainstream. The multi-level perspective emphasises the need to broaden focus past the technology to the co-evolution of technology and society, including not only devices, entrepreneurs and users but the cultural, infrastructural, financial, regulatory, political and other networks that make up the whole picture \cite{rip_technological_1998}. For V2G, the quality of niche development can be evaluated in terms of the quality of first- and second-order learning that is occurring and the quality of institutional embedding, which is an argument for greater interplay, reflexivity and networking among those involved in niche activities and more broadly \cite{hoogma_experimenting_2002}. Supportive stakeholders outside the niche can look for opportunities to embed the niche-innovation into their own practices, for example by learning new sets of engineering heuristics or calling for V2G in design briefs, building and planning codes, or green ratings schemes.

There is significant work underway to further map the V2G landscape and prove benefits, quantify and resolve risks, and normalise grid participation. These trials and demonstrations are a key part of increasing the benefits, reducing the barriers and building compelling narratives around V2G.

\section{Acknowledgements}

This work was supported by the Australian Renewable Energy Agency Advancing Renewables Program under grant 2018/ARP134.

\section{CRediT authorship contribution statement }

Bj{\"o}rn C. P. Sturmberg: Conceptualization, Investigation, Writing - original draft, Project administration, Funding acquisition. 
Laura Jones: Investigation, Writing - original draft.
Kathryn Lucas-Healey: Investigation, Writing - original draft.
Monirul Islam: Investigation, Writing - review \& editing.
Hugo Temby: Investigation, Writing - review \& editing.

\section{Additional information}

\textbf{Competing interests} The authors declare that they have no known competing financial interests or personal relationships that could have appeared to influence the work reported in this paper.

\bibliography{main}

\end{document}